\documentclass[a4paper,5p, sort&compress, preprint]{elsarticle}

\usepackage{graphicx}
\usepackage{amssymb}
\usepackage{float}

\journal{NIM}

\begin{document}

\begin{frontmatter}

\title{
High-resolution ion pulse ionization chamber with \\ air filling for the $^{222}$Rn decays detection}

\author[a1]{Yu.M.~Gavrilyuk}
\author[a1]{A.M.~Gangapshev}
\author[a1]{A.M.~Gezhaev}
\author[a1]{R.A.~Etezov}
\author[a1]{V.V.~Kazalov}
\author[a1]{V.V.~Kuzminov}
\author[a2]{S.I.~Panasenko}
\author[a2]{S.S.~Ratkevich \corref{cor}}
\cortext[cor]{Corresponding author, Tel +380 577075185; +7 8663875149.}
\ead{$ssratk@gmail.com$}
\author[a1]{D.A.~Tekueva}
\author[a1]{S.P.~Yakimenko}
\address[a1]{Baksan Neutrino Observatory,Institute for Nuclear Research RAS, 361609, Neutrino, Russia}
\address[a2]{V.N.Karazin Kharkiv National University, 61022, Kharkiv, Ukraine}

\date{}

\begin{abstract}
The construction and characteristics of the cylindrical ion pulse ionization chamber (CIPIC) with a working volume of 3.2 L are described.
The chamber is intended to register $\alpha$-particles from the $^{222}$Rn and its daughter's decays in the filled air sample.
The detector is less sensitive to electromagnetic pick-ups and mechanical noises.
The digital pulse processing method is proposed to improve the energy resolution of the ion pulse ionization chamber.
An energy resolution of 1.6\%
has been achieved for the 5.49 MeV $\alpha$-line.
The dependence of the energy resolution on high voltage and working media pressure has been investigated and the results are presented.

\end{abstract}


\end{frontmatter}

\section{Introduction}

The problem of detection and elimination of the radioactive background volatile component, which is produced by decays of the $^{222}$Rn and its progenies in the environmental air, takes a noticeable place in low background experiments. Radon, $^{222}$Rn, is an intermediate volatile radioactive isotope of the $^{238}$U chain and the mother isotope of the end part of the following chain: $^{222}$Rn
(
$^{222}$Rn (T$_{1/2}=3.823$ days, $\alpha$-decay, $E_{\alpha}= 5490$ keV)$ \rightarrow
^{218}$Po (T$_{1/2} = 3.11$ min, $\alpha$, $E_{\alpha} = 6003$ keV) $ \rightarrow
^{214}$Pb (T$_{1/2} = 26.8$ min, $\beta$) $ \rightarrow $ $
^{214}$Bi (T$_{1/2} = 19.9$ min, $\beta$) $\rightarrow $ $
^{214}$Po (T$_{1/2} = 164.3$ $\mu$s, $\alpha$, $E_{\alpha} = 7687$ keV) $\rightarrow $ $
^{210}$Pb (T$_{1/2} = 21.8$ y, $\beta$) $\rightarrow $ $
^{210}$Bi (T$_{1/2} = 5.01$ days, $\beta$) $ \rightarrow $ $
^{210}$Po (T$_{1/2} = 138.4$ days, $\alpha$, $E_{\alpha} = 5297$ keV) $ \rightarrow $ $^{206}$Pb (stable) \cite{nudat}
).

A detector of the radon activity is a central element of a control system of environmental air background characteristics.
Air seems to be the optimum choice for the working medium of a radon activity detector.
It should have a large enough volume to support a high sensitivity and a good energy resolution to distinguish between $\alpha$-peaks
from decays of the $^{222}$Rn, $^{218}$Po and $^{214}$Po simultaneously presented in an air test sample.
Due to the low fluorescence efficiency $\eta$,  whose value in the air is $\eta=8.1\cdot10^{-5}$ under the normal pressure, it seems to be unreasonable to build a detector on the base of the scintillation technique \cite{Zhemerev}.
Using a technique of gas ionization seems preferable to the scintillation detector.
There is an obstacle in using the traditional ionization chamber:
the absence of free electron conductivity in the air.
The electrons are captured by electronegative oxygen
molecules within $\sim3\cdot10^{-7}$ s, and positive and negative ions become charge carriers.
The drift velocity of the negative ions in the air is a thousand times smaller than that of the electrons in a gas with electron conductivity.
As a result, the ionization charge collection time becomes of the order of milliseconds.
For example, in the electric field of 200 V$\cdot$cm$^{-1}$, it takes $\sim2.8$ ms for a negative ion to pass a distance
of 1 cm in the air at normal pressure. In order to amplify and register such signals, one needs some nontrivial electronics.
The construction of the detector demands some additional improvements as well.
In particular, the frequency spectrum of $\alpha$-pulses falls into the region of the environmental mechanical noise
and 50 Hz electromagnetic pick-ups. Such factors can create electric noise pulses in the signal circuits of the chamber
due to the microphone effect and/or insufficient screening. Therefore, to prevent such a noise, one needs some additional
arrangements in the electrodes system shielding compared with a traditional ion pulse ionization chamber.

It should be taken into account that most of the pulses recorded in the air ion-pulse ionization chamber are produced due to the $\alpha$-decays of $^{222}$Rn in the working gas  and its daughter nuclei, $^{218}$Po and $^{214}$Po, at the high voltage and collector electrodes. Amplitudes of the 6.00 and 7.69 MeV peaks correspond to $\alpha$-particle energies. The right-hand slope of the 7.69 MeV $\alpha$-peak extends to higher energies due to the pile-up of the energy release from $\beta$-particles generated in the $^{214}$Bi nuclei decay. In determining the 5.49 MeV $\alpha$-peak position in an amplitude spectrum, one should also take into account an additional contribution of the nuclear recoil energy (101 keV) which is not fully transferred into the gas ionization.

A multi-wire ion pulse chamber, with total and fiducial volumes of 25 L and 16 L respectively, was proposed as such a detector in \cite{Kuzminov03}. In order to decrease the total time of the ionization charge collection, the fiducial volume was divided by 0.05 mm longitudinal nichrome wires into 100 ($20\times20\times100$ mm) cells. Each cell had a central anode, and all anodes were connected in parallel. The chamber current pulse time duration reached $\sim3$ ms at the air pressure of 0.83 bar and the supplied voltage of ($-1.4$ kV). The best energy resolution, 3.9\%,
for alpha-line of 5.49 MeV, was achieved under the special low noise conditions. During the long time measurement in the standard underground low-background laboratory, the average energy resolution was $\sim5$\%
\cite{Gavr11}.
For $10^3$~s of measurement at the volumetric radon activity in the air of 10 Bq$\cdot$m$^{-3}$, the high sensitivity of the detector allowed us to achieve the statistical uncertainty better than 10\%.
However, this designed sample of the detector had a large enough mass, $\sim150$ kg, which, together with the increased sensitivity to a noise, made it difficult to be used widely.

A possibility to considerably improve the characteristics of the radon ion-pulse chamber was found during the study of characteristics of the ion-pulse chamber intended for the measurement of the surface $\alpha$-activity of different materials \cite{Gavr09}.The chamber consisted of two vertical cylindrical sections of 91 mm in diameter, 74 mm high, and of 0.48 L volume each, placed one over the other and separated with a common high-voltage mesh electrode. A sample under study was placed at the bottom of the lower section. The bottom served as the charge collector electrode. The upper section protected the fiducial volume of the lower section from $\alpha$-particles emitted by the surface of the upper charge collector electrode, and also from alpha-particles of radon and its daughter nuclei decay of the working gas. Nitrogen or air were used as a working gas. The chamber current pulse time duration reached $\sim15$ ms at the 0.83 bar air pressure and ($-2$ kV) high voltage.

An average resolution of the 5.49 MeV $\alpha$-peak measured in the ground level laboratory during long-time measurements was found to be 4.3\%.
This value was obtained by means of a reduction to the infinite preamplifier discharge time of pulse shapes recorded with a digital oscilloscope and by summation of the lower section pulse with the corresponding noise signal of the upper section. The amplitude of the noise in the summed pulse has been thus lowered because the noise components are in opposite phase. The result was achieved without any special low noise shielding to eliminate the microphone effect.

The obtained data showed a trend for further improvement of characteristics  of the ion-pulse chamber intended for the spectrometric measurements of the radon and it's daughters $\alpha$-decays in the air. A construction of the cylindrical ion pulse ionization chamber (CIPIC) was designed. CIPIC was produced and its characteristics were measured. The results of this work are presented below.
\begin{figure} [pt]
\begin{center}
\includegraphics*[width=1.75 in,angle=270.]{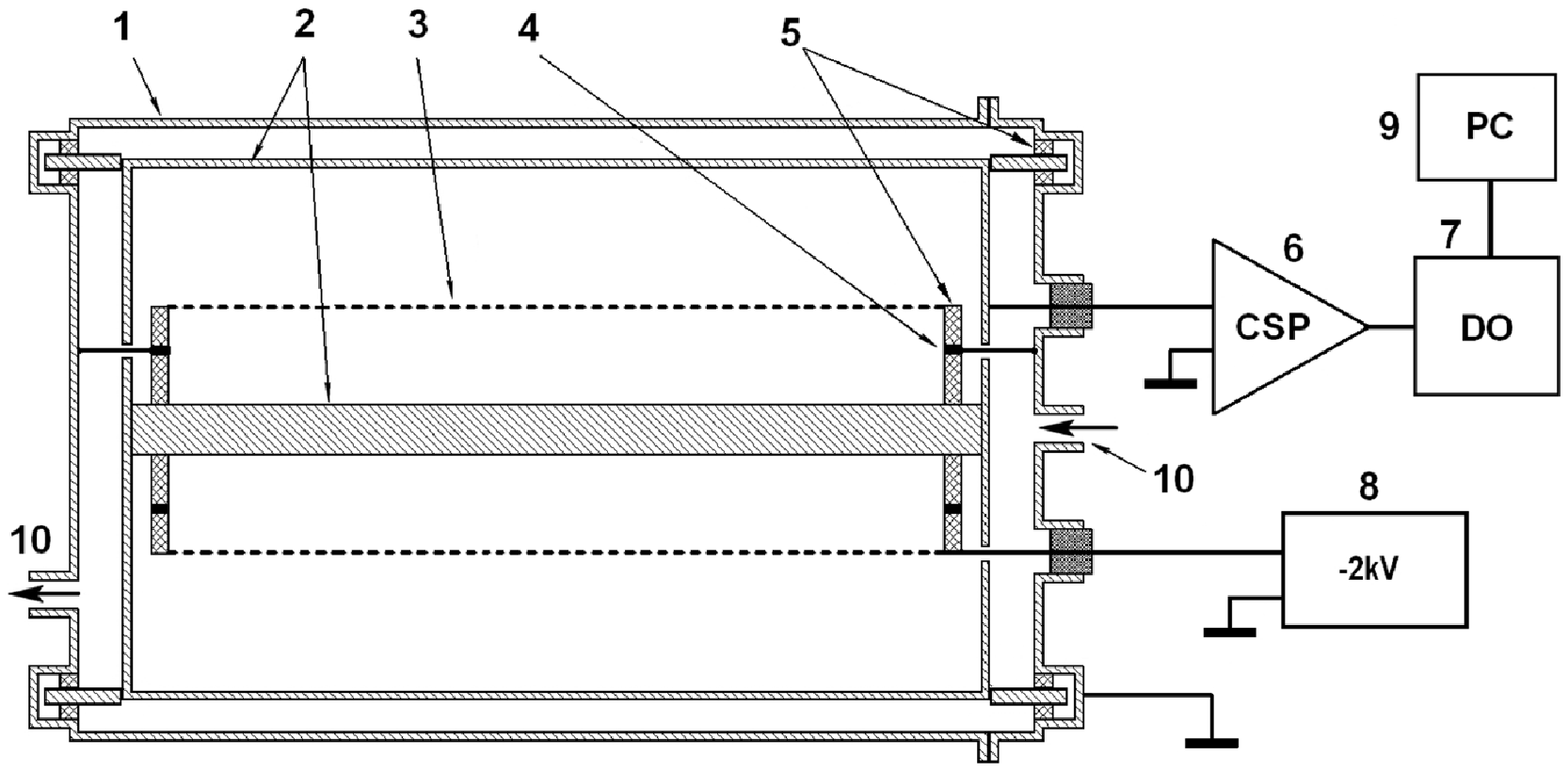}
\caption{\label{fig1} Schematic longitudinal cross-section and electrical circuit of CIPIC.
1. Camera body. 2. Collector electrode (anode).
3. High-voltage multi-wire greed electrode (cathode).
4. Guard electrode.
5. Insulators.
6. Charge-sensitive preamplifier (CSP).
7. Digital oscilloscope (DO).
8. High-voltage power supply.
9. Computer (PC).
10. Inlet and outlet chamber blow nipples.
}
\end{center}
\end{figure}

\section{CIPIC construction}

In considering the origin of the microphone effect, it was assumed that an electrical signal at the output of the charge sensitive preamplifier (CSP) could appear as a result of the various processes activated by mechanical shaking of the chamber body. These could be variations in the capacity of the high-voltage electrode due to its relative shift with regard to the collector one, or it could be oscillation of the output signal wire in the electric field of the supplying high-voltage wire or vibrations of the massive parts of CSP and high-voltage circuit. An option to automatically compensate the capacity variations was chosen to exclude the first process. The result was achieved by placing the high-voltage electrode between the two plates of the collector electrode. The signal and high-voltage electrodes should be as short as possible, well fixed inside the chamber, and screened from each other to minimize the effect of the second process. The screening of these electrodes serves also to eliminate signals from the collection of charge ionization produced by $\alpha$-particles outside of the chamber. The third process could be avoided by a proper assemblage of the CSP parts, its safe mounting in the electronics compartment, and by screening it well from the high-voltage circuits. Taking all the above into consideration, the variant of a cylindrical chamber with coaxially mounted electrodes has been chosen.

A schematic longitudinal cross-sectional view and the electrical circuit of CIPIC are shown in Fig.\ref{fig1}.
The anode collector electrode (2) consists of the inner and outer cylinders connected by the end discs. The cathode high-voltage grid electrode made of tungsten wire (3) is mounted on the central cylinder by means of the Teflon disk insulators (5), which have the guard electrodes (4). The electrode system of the chamber is mounted in the pressurized cylindrical body (1) having inlet and outlet blow nipples (10). The fiducial volume of CIPIC is 3220 cm$^3$. The length of the chamber is 45 cm and its mass is $\sim6$ kg (including the electronics compartment).

The chamber was horizontally suspended inside of a metallic computer carcass by means of silastic guy cable. A negative high-voltage of ($-1.5$ keV) was supplied from MANTIGORA HV-2000N (8) through a RCRC-filter. Signals were taken off with a custom CSP (6). Both, the high-voltage filter and CSP, were placed in the electronics compartment and screened from each other by using metallic walls. Pulses from the CSP output were fed to the input of the digital oscilloscope (DO) LA-n10-12USB (7) and then out from the DO output ($V[i]$, $i=1,...,i_{max}$) to the USB-port of a PC (9). PC manages the DO modes and also records digitized pulses. In the series of the test measurements, the sampling rate was chosen to be 1.56 MHz. DO starts recording signals when the amplitude of pulse exceeds a specified threshold. ``Prehistory'' (a segment of a noise line that precedes the pulse) and ``history'' (the pulse itself) are present in the recorded frame.

\section{Working characteristics of the CIPIC}

\begin{figure}
\begin{center}
\includegraphics*[width=2.25in,angle=0.]{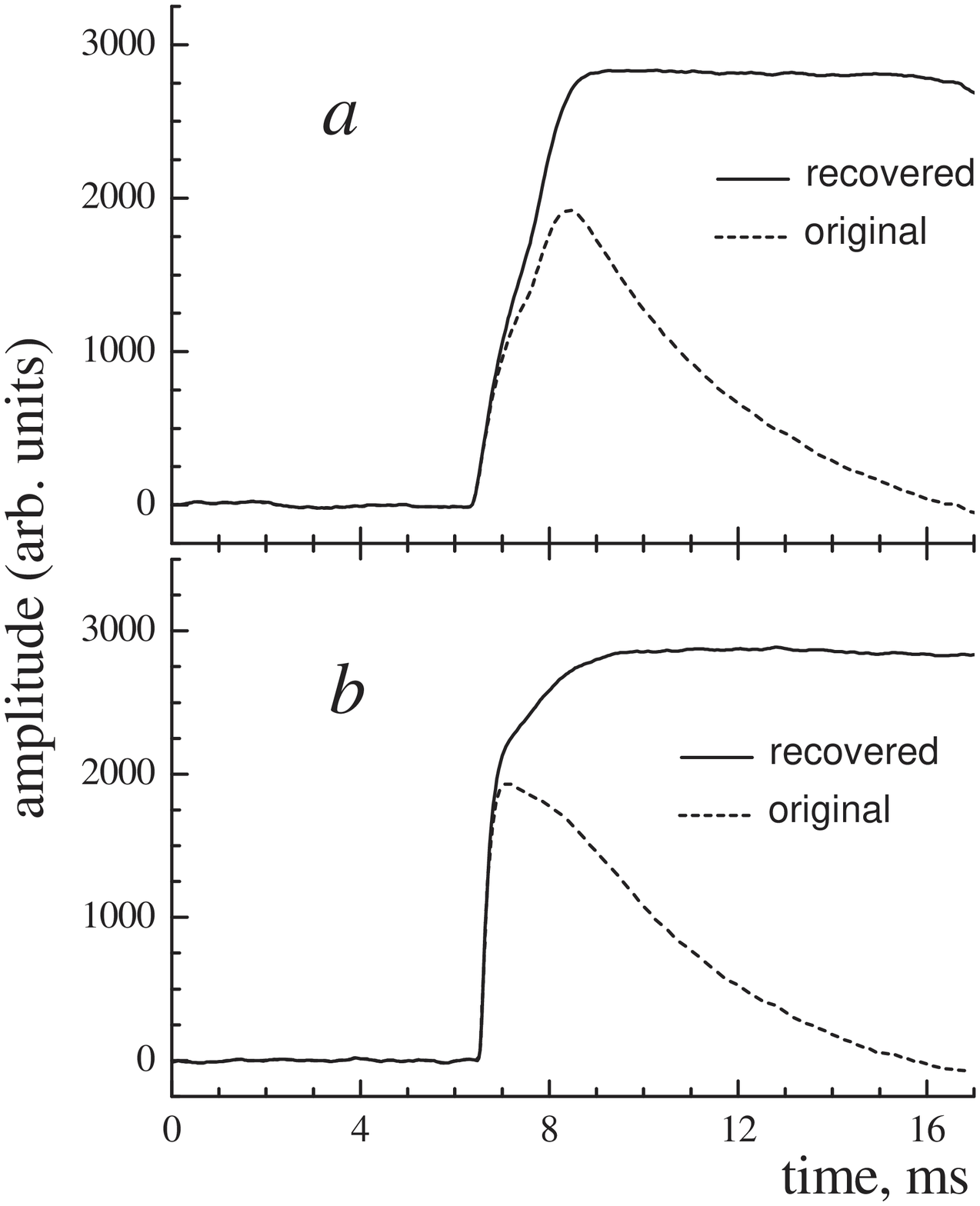}
\caption{\label{fig2} Examples of pulse from the $\alpha$ - particles with an energy of 7.69 MeV:
dashed curve - original pulse of CSP; dark curve - recovered the charge pulse. \emph{a} - camera with air;
\emph{b} - camera with nitrogen.}
\end{center}
\end{figure}

CIPIC was filled with an air sample taken from the laboratory air using an aquarium air pump.
Atmospheric pressure at the height of the Baksan Neutrino Observatory (1750 m) is 620 Torr ($\sim0.83$ bar).
A pulse recorded from the CSP output to the PC memory is shown in Fig.\ref{fig2}\emph{a} (dashed curve).
CSP was built to satisfy the minimum noise requirement, and it had an additional amplifier
incorporated into it whose gain was $\sim400$.
The decay constant of the CSP self-discharge ($\tau_1$) was equal to $\sim4$ ms and was not optimized
in accordance with the ratio of $\tau_1/\tau_2\approx10$ in comparison with a pulse current duration ($\tau_2$) equal to $\sim3$ ms.
It is difficult to fulfill this condition due to the fact that $\tau_2$  depends on various
operating conditions of the camera (power supply voltage, air pressure, etc...).
The amplitude value at the maximum of deposited pulses may vary for the same  energy since this value for a finite CSP discharge constant is a function of the current density.
The true value of the charge produced by an $\alpha$-particle can be reconstructed by introducing a correction
for the CSP self-discharge via conversion to the infinite discharge constant.
If this correction is introduced, the amplitude value is determined by averaging over the prescribed number of points from
the next time interval to the instant when the current pulse ceases.

The proposed digital pulse processing is effective at minimizing the problems and is a
simple method without complex electronics used by other previous techniques.
It could be applied in the online or offline modes.

\subsection{Digital pulse processing}

\emph{At first}, the digitized primary pulse was smoothed by ``moving average on several points'' \cite{Savitzky64}, which is defined as
\[V_i^{av}=\frac{1}{\bigtriangleup n}\sum\limits_{j = i}^{j=i+\bigtriangleup n}V^j,\]
where $i=1,...,\{i_{max}-\bigtriangleup n\}$, $i_{max}$ - the last time channel of the pulse.
Since this filter takes an average value for a data length $\bigtriangleup n$ ($\bigtriangleup n \ll i_{max}$), the high frequency noise component is filtered out and the output is much smoother than the original data.

\emph{Then}, the task of processing the pulse shape is to restore the value of the total charge of the ionization, which corresponds to the amplitude of the output of the amplifier, which has no self-discharge.
An input CSP current pulse of any duration can be expressed as a sum of non-separable shorter currents.
This processing is based on the assumption that a CSP's response function for a pulse of long duration is equal to the sum of the response functions of its shorter components (pulses).

The response function (r.f.) can be defined in different ways:
1) in an analytical form if you know the law of transformation of short current pulses;
2) as an empirical expression that best approximates the shape of the measured output pulse at the input of a short pulse;
3) a table of values of the output pulse for a short input pulse.
As a rule, it is difficult to obtain the overall analytical form of the r.f of a device that combines the  input charge-sensitive amplifier and a subsequent amplifier with high gain. For this reason, the description r.f. preamplifier was used option (2).
An exponential function $f(t) = exp(-t/\tau)$ be chosen to account for the CSP self-discharge.
This can possible if the time interval between the top pulse and the point of transition recession through zero considerably longer than the pulse rise time.
It is assumed that the amplitude of a digitized pulse is not distorted by CSP in the first time digit channel and is equal to the current at that time step.
The corrected response to the self-discharge in each digital channel ($i=1,...,i_{max}$) of the CSP output pulse
can be calculated using the recursive algorithm:
\[V_i^{rec} = \left\{ {\begin{array}{*{20}{c}}
{V_i^{av},\begin{array}{*{20}{c}}
{}&{i = 1,}
\end{array}}\\
{V_{i - 1}^{rec} + V_i^{av} - V_{i - 1}^{av} \times \exp ( - {{\Delta t} \mathord{\left/
 {\vphantom {{\Delta t} \tau }} \right.
 \kern-\nulldelimiterspace} \tau }),  i > 1.}
\end{array}} \right.\]

The width of the time channel $\Delta t$ is equal to 0.64 $\mu$s.
The value of $\tau$ was chosen to be $\tau=3.296$ ms.
The corrected total ionization charge pulse $V^{rec}(t_i)$ is shown in the Fig.\ref{fig2} (dark curve).

\subsection{Spectrometric properties}

\begin{figure}
\begin{center}
\includegraphics*[width=2.5in,angle=0.]{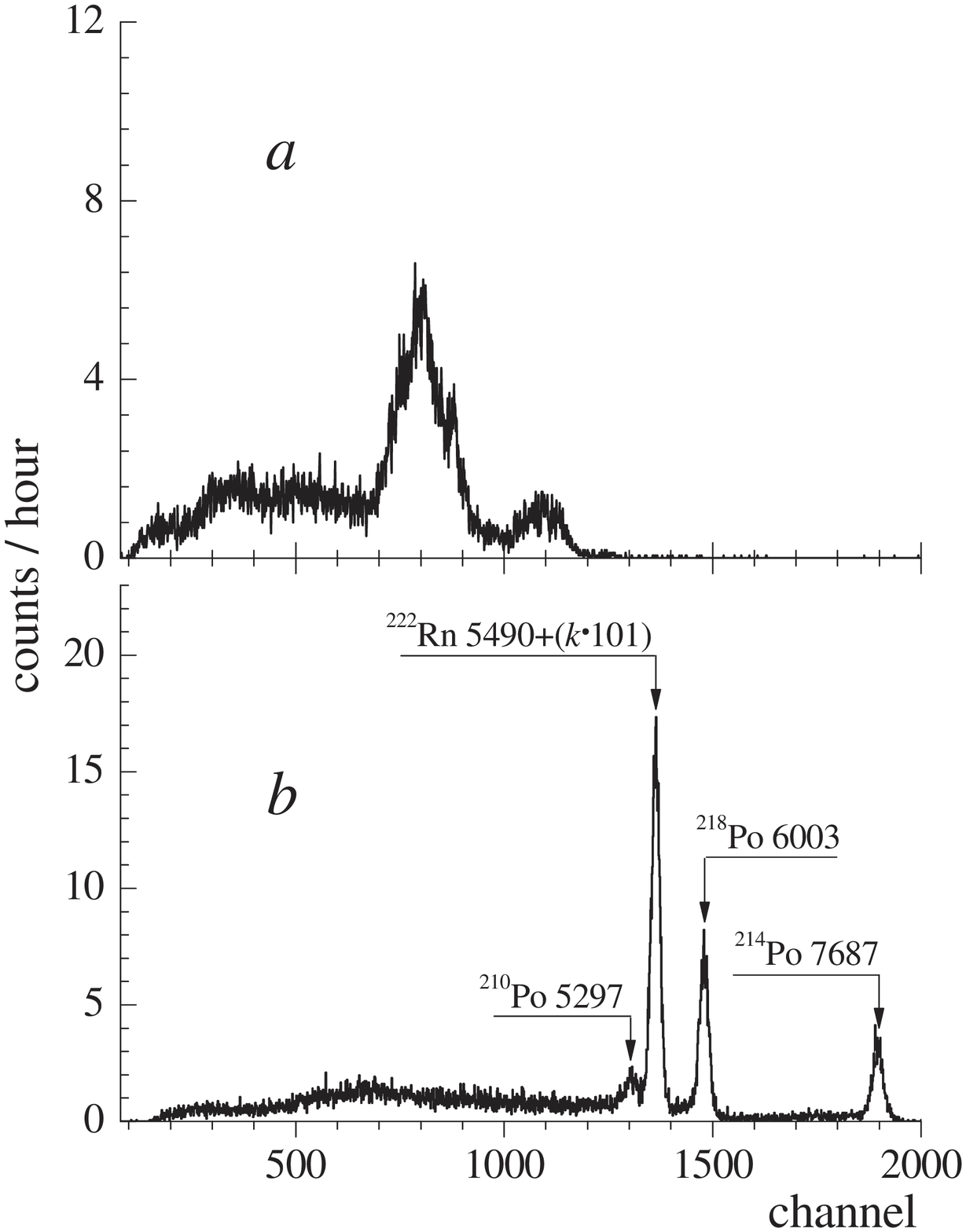}%
\caption{\label{fig3}
Normalized to 1 h height spectrums of $\alpha$-particles from the radon and it's daughter decays registered the CIPIC, filled with air at 620 mm Hg (0.83 bar) and voltage (-1.5 kV): \emph{a}) the height spectrum of the initial pulses; \emph{b}) the height spectrum of the reconstructed pulses.
The numbers indicate the energy of the $\alpha$-peaks in keV.
}
\end{center}
\end{figure}

An amplitude spectrum, normalized to 1 h, of $\alpha$-particles from radon and its daughter decays collected during a 16.2 h observation process is shown in Fig.\ref{fig3}\emph{a}.
The amplitude values were read in the maximum of the recorded pulses.
As it can be seen from the  figure, the obtained spectrum is of little use for spectrometric measurements.
An amplitude spectrum constructed from the same pulses after their recovering is shown in Fig.\ref{fig3}\emph{b}.

\begin{figure}
\begin{center}
\includegraphics*[width=1.5in,angle=0.]{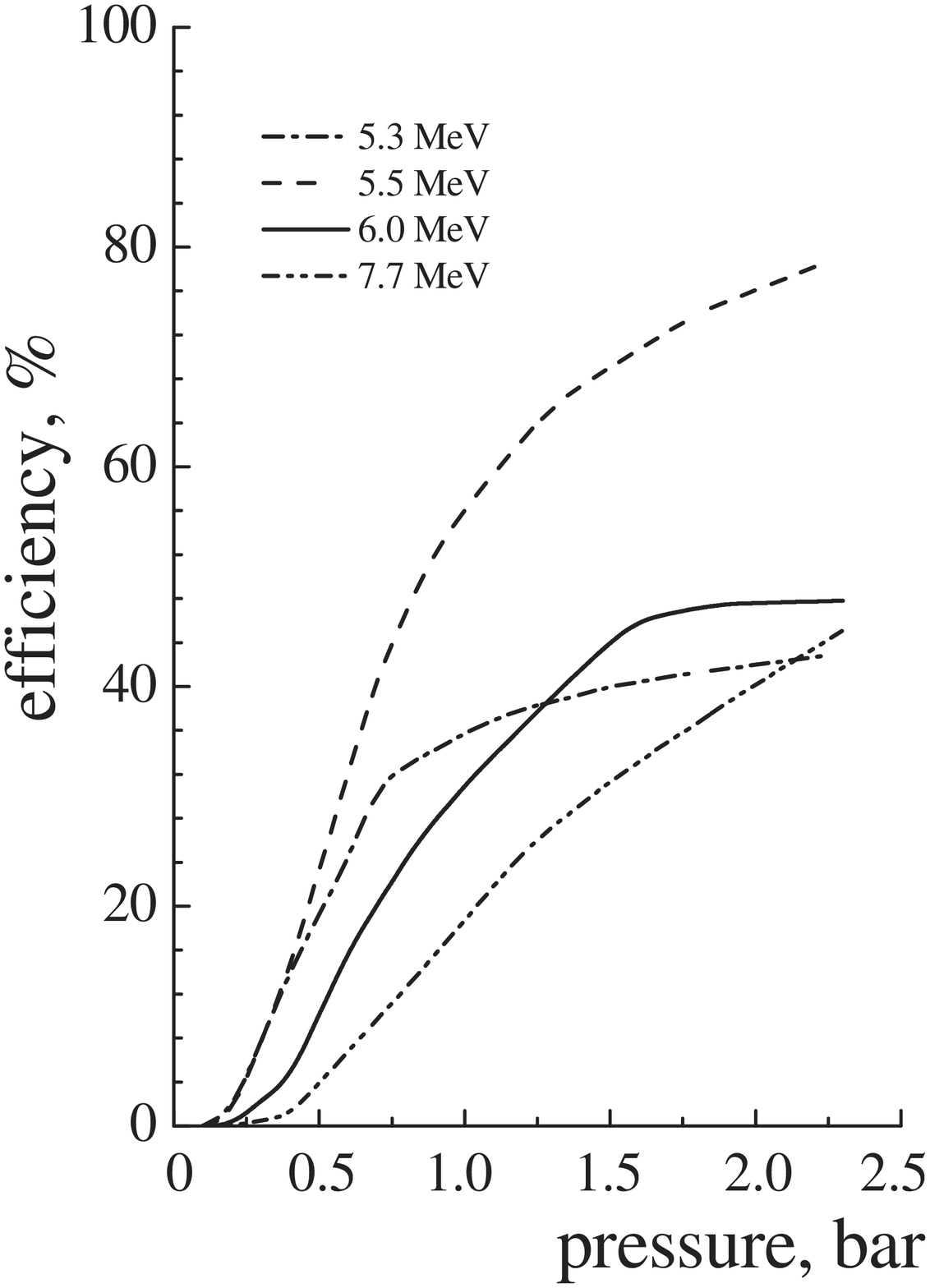}%
\caption{\label{fig4}
The calculated dependence of the complete energy absorption efficiency from the air pressure in the chamber for four $\alpha$-lines.
}
\end{center}
\end{figure}

The $\alpha$-peaks of the $^{210}$Po (5297 keV - \emph{a}), $^{222}$Rn ($5490+k \cdot 101$ keV - \emph{b}), $^{218}$Po (6003 keV - \emph{c}) and $^{214}$Po (7687 keV - \emph{d}) are clearly visible on the spectrum.
The ratio, $a:b:c:d$, of the peak areas is equal to $0.075:1.00:0.52:0.26$. The source of $^{210}$Po is located on the surface of the electrodes in the form of the $^{210}$Pb parent atoms which have fallen on the material, mostly from the environment, since its manufacturing. The activities of $^{210}$Po and $^{222}$Rn are far from equilibrium due to this fact. On the other hand, the activities of $^{222}$Rn, $^{218}$Po and $^{214}$Po should be in equilibrium but the areas of the corresponding peaks differ considerably. As was already mentioned, radon decay occurs in the air, and therefore a peak of Rn includes all $\alpha$-particles except for those whose trajectories have left the fiducial volume of the detector.
These ones release only a part of their energy into the gas.
Moreover, these events form a low energy flat part of the spectrum. The daughter radon decay products are positively charged in 90\%
and negatively charged in 10\%
of release cases \cite{Kotrappa88}. That is why they are deposited asymmetrically on the surfaces of the negative grid and positive collecting electrode. Alpha-particle tracks from the decays on the surface of the material are directed inward in half of the cases. The other half goes into the gas. A part of the tracks enters the wall. These tracks do not contribute to the peak formation.
The energy absorption efficiency was calculated  via a Monte Carlo simulation of trajectories of alpha-particles on the basis of the
GEANT code \cite{Agostinelli2003}.
The calculated dependence of the complete energy absorption efficiency, $\varepsilon_\alpha$, on the air pressure for these four $\alpha$-particles is shown in Fig.\ref{fig4}.
A ratio of $\varepsilon_\alpha$-values for $\alpha$-particles from $^{222}$Rn, $^{218}$Po and $^{214}$Po, $b:c:d$, is equal to $1.00:0.52:0.27$ at 620 Torr, which is in a good agreement with the $\alpha$-peak areas ratio mentioned above.

\begin{figure}
\begin{center}
\includegraphics*[width=2.15in,angle=270.]{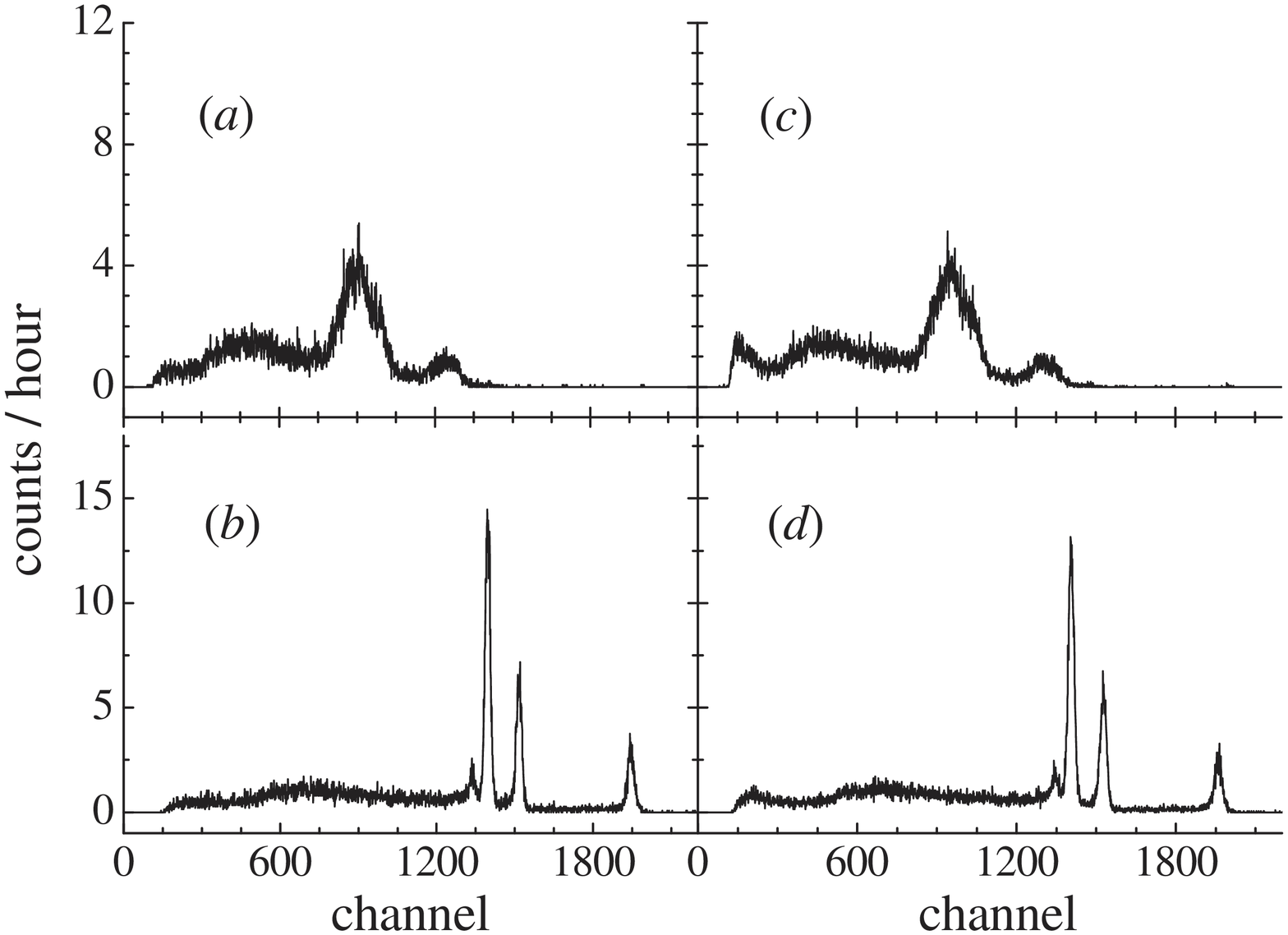}%
\caption{\label{fig6}
The pulse height spectra from the decay of radon and its daughters decays in CIPIC
filled with air to 620 mmHg at a voltage $-1.75$ kV (
\emph{a} - the height of the direct pulse and \emph{b} - the height of the recovered pulse)
and voltage correspondingly at $-2.0$ kV (\emph{c} and \emph{d} respectively.)
}
\end{center}
\end{figure}

\begin{figure}
\begin{center}
\includegraphics*[width=2.5in,angle=270.]{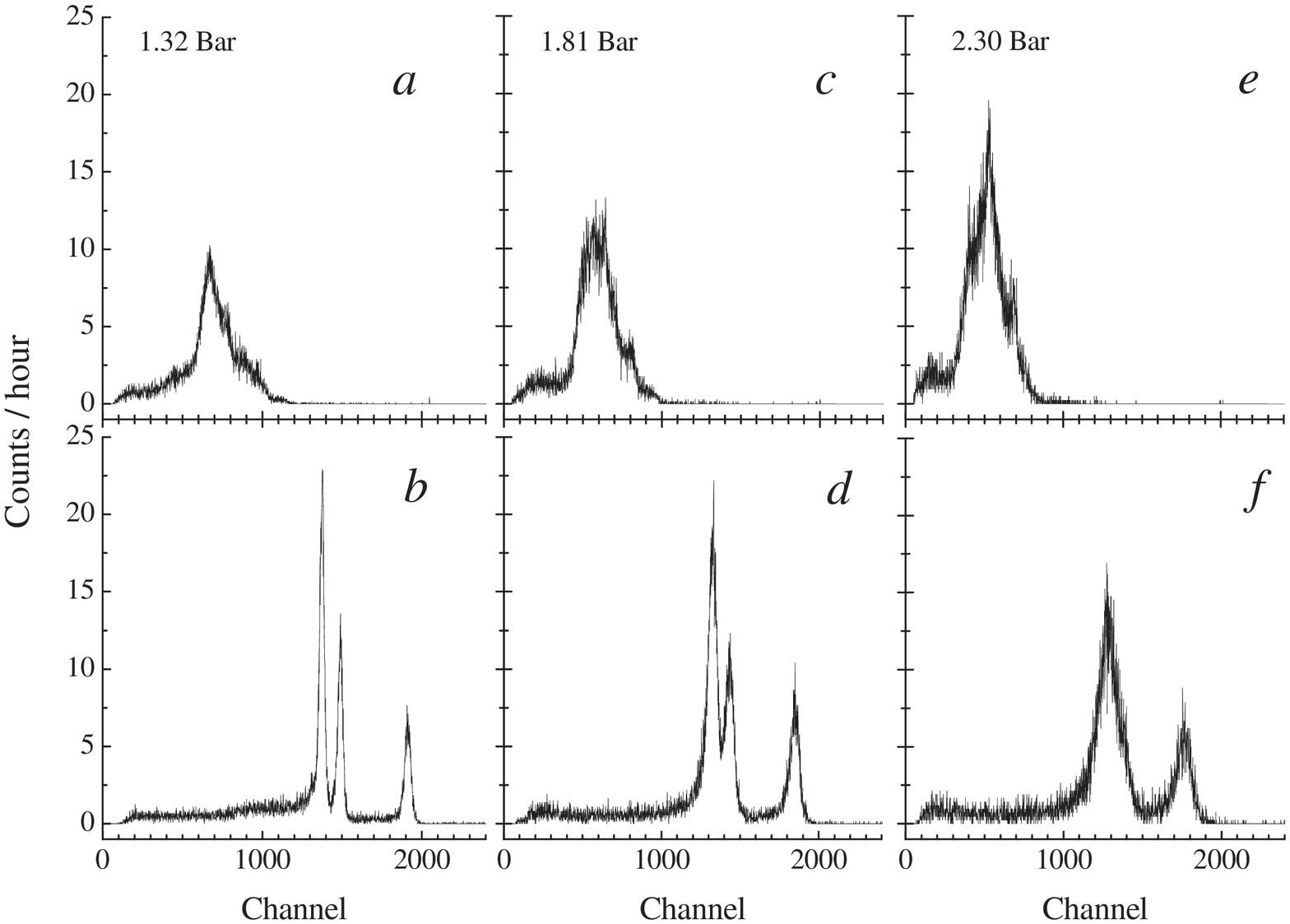}%
\caption{\label{fig7} The pulse height spectra from the decay of radon and it's daughters decays in CIPIC
filled with air at a voltage $-1.5$ kV: (\emph{a} - the height of the  direct pulse and \emph{b} - the height of the recovered  pulse) at 1.32 bar, (\emph{c} and \emph{d}) at 1.81 bar and (\emph{c} and \emph{d}) at 2.30 bar, correspondingly).}
\end{center}
\end{figure}

The energy resolution of the $^{222}$Rn $\alpha$-peak is equal to $(1.7\pm0.1)$\%.
A total energy of this peak defined from the calibration with the $^{210}$Po, $^{214}$Po and $^{218}$Po $\alpha$-lines is equal to $5537\pm6$ keV. This means that the recoil nuclear adds $(47\pm6)$ keV to the $\alpha$-particle energy. A value of this addition depends on several factors, including the following: a) the total portion of kinetic energy that was transmitted into ionization by the slow recoil nuclear, b) the part of an ionization electron charge recombination on its own ions \cite{Lindhard63}, and c) the part of the charge recombination in the ionization column during the time of charge separation under the influence of the electric field \cite{Raizer80}. The latter component depends on the electric field value. To find this dependence, the measurements were repeated at the high voltage of ($-1.75$ kV) and ($-2.0$ kV). The corresponding spectra normalized to 1 h are shown in Figs.\ref{fig6} and \ref{fig7}. The energy resolution was equal to $(1.8\pm0.1)$\%
in both cases.

The peak has shifted from the $1365\pm1$ channel ($-1.5$ kV) to the $1397\pm1$ channel ($-1.75$ kV), and to the $1406\pm1$ channel ($-2.0$ kV). The total energy of each peak was found to be $5532\pm6$ keV ($-1.75$ kV) and $5525\pm6$ keV ($-2.0$ kV). It follows that a portion of $^{218}$Po nuclear recoil energy in the $^{222}$Rn $\alpha$-peak does not depend, within $1\sigma$ limits, on the high voltage. It is possible to estimate the averaged coefficient $k$, the portion of $^{218}$Po nuclear recoil energy that has passed into ionization as $k=(41\pm6)/101=0.41\pm0.06$, if one neglects the loss of primary ionization charge in the process of recombination on its own ions.

\begin{figure}
\begin{center}
\includegraphics*[width=2.15in,angle=0.]{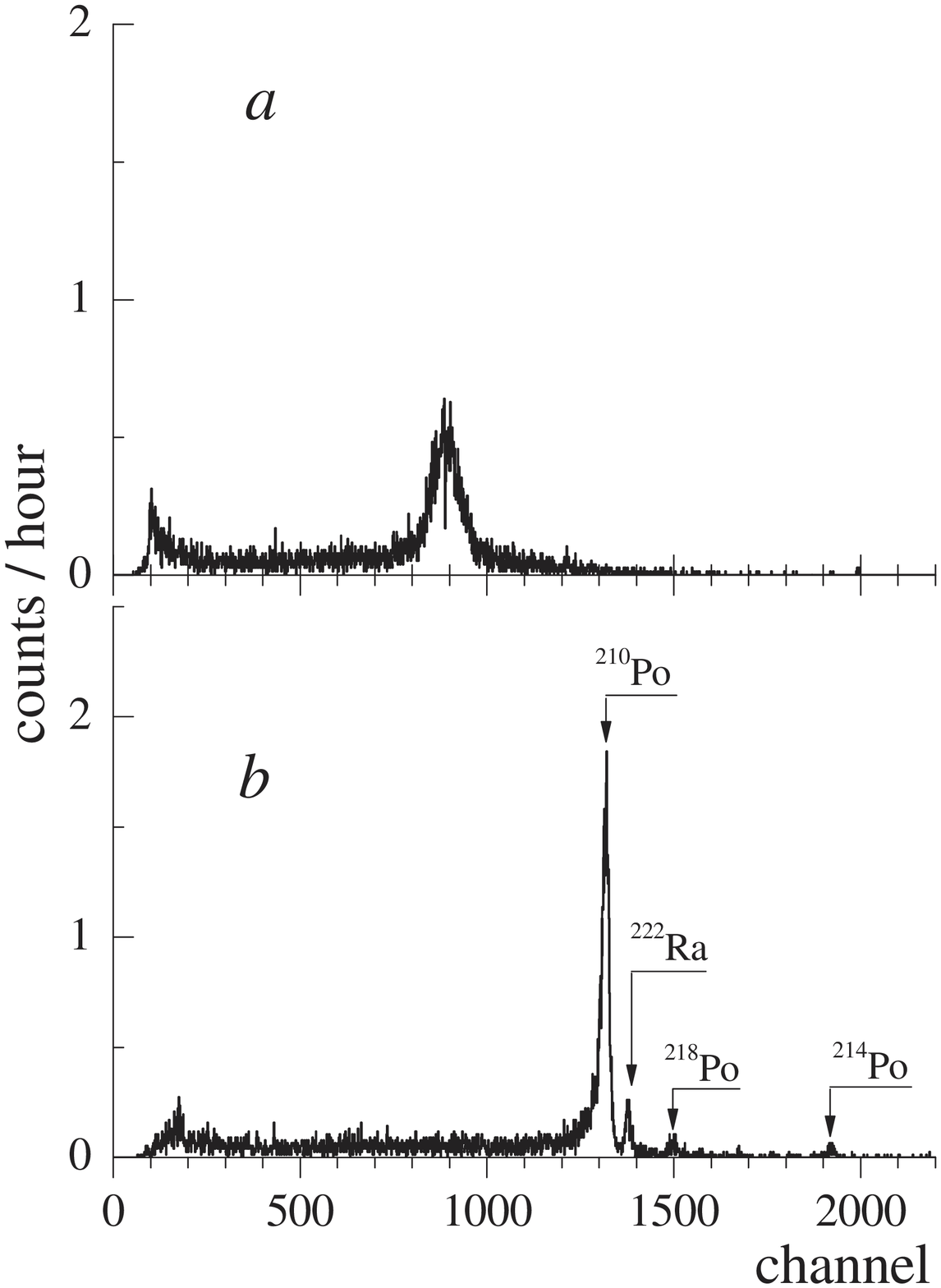}%
\caption{\label{fig5} The pulse height spectra from the decay of radon and it's daughters decays in CIPIC
nitrogen filled to 620 mm Hg, with a voltage $-2.0$ kV
(\emph{a} - the height of the direct pulse and \emph{b} - the height of the recovered pulse).}
\end{center}
\end{figure}

The ultimate sensitivity of CIPIC to an air radon activity measurement could be estimated if its own background is known. CIPIC was connected to a Dewar with a silicon rubber tube to blow off the chamber with liquid nitrogen vapor. Nitrogen was directly deflated into  the atmosphere with a volume velocity of 10 L$\cdot$h$^{-1}$.A parallel measurement of $\alpha$-background was started after 20 h of this blowing procedure and continued for 76.5 h at the ($-2.0$ kV) high voltage. The pure nitrogen is not an electronegative gas, and the electrons are the carriers of the negative charge. The electrons and negative ions could simultaneously be the carriers of the negative charge if oxygen is present in the nitrogen gas. The shape of the CSP output pulse for the chamber filled with nitrogen is
shown in Fig.\ref{fig2}\emph{b} (the dashed curve - original pulse and that for the recovered one is the dark curve).
The collected direct (\emph{a}) and recovered (\emph{b}) spectra, which are normalized to 1~h, are shown in Fig.\ref{fig5}.

The peaks of  $^{210}$Po, $^{222}$Rn, $^{218}$Po and $^{214}$Po are at the 1320, 1379, 1498 and 1921 channels, respectively. A resolution of $^{222}$Rn is $(1.1 \pm 0.1)$\%.
A total energy release at this peak is $5529 \pm 6$ keV, therefore the contribution of the nuclear recoil is $39 \pm 6$ keV. This value is equal to the one obtained for the air within statistical uncertainties.

An energy resolution for the 5297 keV peak was calculated for its right slope and for the whole peak width are $(1.1\pm0.1)$\%
and $(1.7\pm0.1)$\%,
respectively. The fact that the peak extends to the lower energy indicates that $\alpha$-particles lose small amount of their energy in the matter. This could be due to a location of parent nuclei in the surface relief irregularities of the unpolished electrodes. A calculated surface $\alpha$-activity $A_S$ of $^{210}$Po is equal to $A_S=(10.0\pm0.3)$ $\alpha\cdot$(h$\cdot100$ cm$^2$)$^{-1}$. A count rate of the $^{222}$Rn decays under the peak is equal to $(3.0\pm0.2)$ $\alpha\cdot$h$^{-1}$. It gives a radon volume activity $A_V=(0.54\pm0.04)$ Bq$\cdot$m$^{-3}$ taking into account the value of $\varepsilon_\alpha=0.48$. There could be the following possible sources of radon: the Dewar inner surface, the liquid nitrogen itself, the rubber tubes or the chamber inner materials. The blowing off rubber tubes were pinched after this measurement and the new measurement was made during 64.3 h with the isolated chamber. This allowed us to obtain the velocity $\Lambda$ of the radon emission from the inner chamber material and closed parts of the rubber tubes. The value is equal to $\Lambda=190\pm6$  atoms$\cdot$h$^{-1}$. It provides an increase in the radon count rate under the peak
of $0.68\pm0.02$ $(\alpha \cdot$h$^{-1})\cdot$h$^{-1}$.

One can see a small peak at channel 175 in Fig.\ref{fig5}\emph{b}.
It was assumed that this peak corresponds to the energy of protons generated in the reaction $^{14}{\rm N} (n,p)^{14}{\rm C}+626$ keV. A cross-section of this reaction for slow neutrons is equal to $\sigma_{n,p}^{th}=(1.83\pm0.07)$ barn \cite{Gledenov93}. The specific rate of the neutron reaction with nitrogen in CIPIC was estimated using the results of the thermal neutron flux measurements \cite{Alekseenko1302, Alekseenko1303} and the value of $\sim 4$ h$^{-1}$ was obtained. A count rate of the events under the peak was found to be equal to $(2.2 \pm 0.3)$ h$^{-1}$. This gives a value $(2.5\pm0.3)$ h$^{-1}$ for the specific rate of the thermal neutron reactions with nitrogen in CIPIC when the efficiency of the proton energy total absorption is 0.87. This result does not contradict with the estimated one.

The $\alpha$-particle registration efficiency increases appreciably with the pressure rise as seen in Fig.\ref{fig4}. The increase in pressure leads to larger volumes of the sample and consequently to the increase in sensitivity of measurements. The construction of CIPIC allows it to be pumped out or filled to the pressure; the chamber working characteristics were tested at 0.83, 1.32, 1.81 and 2.30 bar. The high voltage was ($-1.5$ kV) in all of these cases. The obtained initial and recovered spectra normalized to 1 h are shown in Figs. \ref{fig3} and \ref{fig7}.

The radon peak position in the initial spectra shifts to origin of coordinates with the increase in air pressure due to recombination and the rise of the current pulse duration.

The dependence of the amplitude of the recovered pulses on the air pressure is much smaller but present. The energy resolution of the radon and its daughter $\alpha$-lines become worse with the increase in pressure. Partially, this is connected with the increase of microphone noises induced by relaxation of the chamber body mechanical tension after a pressure change. Time of relaxation increases with the increase in pressure. Therefore, it is difficult to use the pressurized air to increase the sensitivity of the radon monitoring at the frequent periodic sample exchange mode. Nevertheless, the obtained results indicate that the sensitivity of measurements could be better, with good enough energy resolution retained, by using CIPIC of larger working volumes, up to $\sim 10$ L, at the normal air pressure.

\section{Results and discussion}

As follows from the above consideration, the energy resolutions of $(1.7\pm0.1)$\%
and $(1.1\pm0.1)$\%
could be achieved for the $\sim5.49$ MeV $\alpha$-peak in the case of the $^{222}$Rn decays registration with the cylindrical ion pulse ionization chamber filled up to 620 Torr by air or nitrogen, respectively. These good values are provided by the chamber appropriate construction and by the digital method of pulse registration and processing. The method enables us to improve considerably the amplitudes and noise pulse characteristics in comparison with in comparison with the usual analogous method of the pulse registration after the shaping.

It is necessary to compare the obtained energy resolution with the theoretical one to estimate the quality of the result. The total number of ion pairs, $N$, created by an $\alpha$-particle is equal to $N=E_0/w$, where $E_0$ is the $\alpha$-particle energy loss in the gas, and $w$ is the average energy of one ion pair creation. The $w$ is equal to 35.5 eV in air and 36.6 eV in nitrogen gas \cite{Siegban65}. A dispersion $\sigma^2$ of  $N$ is equal to $\sigma^2(N)=F\cdot N$, where $F$ is the Fano factor. A dispersion corresponds to the Poisson one if $F=1$.  The energy resolution, $R$, of an $\alpha$-peak which has a Gaussian shape is defined usually as a ratio of the Gaussian width at half of the peak height to the peak energy and is equal to $R=2.35 \sqrt {F \cdot N}$. The $F$ value for the (Ar+0.8\%CH4)
gas mixture is equal to 0.19 \cite{Grupen96, Fano} and the value of energy resolution for such a gas is better by 2.27 times than the one when $F=1$. The $F$ values for the air and nitrogen are not known and could be taken as $F=1$. The calculated value of $R$ for the $^{222}$Rn peak in the air and nitrogen is 0.60\%.
It can be seen from a comparison that the measured and calculated values differ within the factor of three. The residual noise of CSP is distinguishable in Fig.\ref{fig2}$b$ and an insufficient quality of the recovery procedure could be the possible reason of such a difference. Nevertheless, the achieved level of the energy resolution allows one to determine the Fano factor for the gases used in this study as well as for other gases by comparing the energy resolutions for two $\alpha$-line energies.

The sensitivity of CIPIC allows one to achieve 10\%
of statistical uncertainty during $10^3$ s during the 65 Bq$\cdot$m$^{-3}$ radon activity measurement of the air sample under the 620 Torr pressure. Therefore, the chamber could be used for the radon monitoring of the air in working rooms and living quarters with the averaged radon volume activity of $\sim40$ Bq$\cdot$m$^{-3}$ and in the automatic permanent-cyclical mode with the $\sim 1.2 \cdot 10^3$ s duration of a ``sampling-measurement'' cycle.

Gas emitted by the chamber's inner materials creates during 24 h the radon activity of 4.5 mBq in the CPIC volume.
A number of events under the radon peak from this background source would be 187 at first 24 h of measurement.
This would correspond to 1.8 Bq$\cdot$m$^{-3}$ of initial radon activity of the air sample, at 620 Torr, if one assumes that the background effect is equal to radon activity of the fresh air sample. This data allows one to estimate the ultimate sensitivity of CIPIC which is intended to measure the radon activity in a specially refined air for low background laboratories. The obtained result could be compared with the sensitivity of the radon detector which was designed in \cite{Kiko} on the basis of the electrostatic precipitation of the radon daughters from the 418 L stainless steel vessel on the surface of the surface-barrier semiconductor $\alpha$-particle detector. The radon detection limit of that detector(defined as signal/background=1) was estimated to be 70 $\mu$Bq$\cdot$m$^{-3}$.
The ratio between the sensitivities of the detector described in the present paper to the one described in \cite{Kiko} is $2.6\cdot10^4$.
It could be improved by at least $10^2$ times if the sources of the CIPIC radon background are determined and eliminated. The ceramic material both of the high voltage and signal electrodes or/and inward diffusion of radon through the rubber tubes could be the sources of the background radon. The question needs additional investigation.


\section{Conclusion}

The possibility to achieve the energy resolution comparable with the theoretical one in the spectrometric measurements is shown in this publication. The work is based on the experimental measurements fulfilled with CIPIC, a newly designed version of the ion pulse ionization chamber.

The analysis of the shape and front time duration of $\alpha$-particle pulses allows one, in principal, to determine the drift velocities of positive and negative ions in any electronegative gases. It is worth mentioning here that such information is practically absent in the reference books.

An ion charge collection chamber could be used for measurements with any chemically neutral gases and their mixtures. CIPIC, in particular, could be filled with the BF$_3$ gas up to the pressure of $\sim1$ bar for the high efficiency registration of the thermal neutrons. The energy resolution of $\sim5$\%
is expected in the region of the reaction $^{10}{\rm B}(n,\alpha)^7{\rm Li}+2.761$ MeV energy release.

The work was made in accordance with INR RAS and V.N.Karazin KhNU  plans of the Research and Developments.

\newpage


\begin{thebibliography}{5}

\bibitem{nudat}
Nuclear structure \& decay Data (BNL, National Nuclear Data Center): http://www.nndc.bnl.gov/nudat2/.

\bibitem{Zhemerev}
A.V.~Zhemerev and B.M.~Stepanov, 
\emph{The Physics of Pulsed Radiation-Induced Excitation of Air Glow} [in Russian], Energoatomizdat, Moscow.1986.

\bibitem[{Kuzminov (2003) Kuzminov}]{Kuzminov03}
V.V.~Kuzminov,
\emph{Ion-pulse Ionization Chamber for Direct measurement of a Radon Concentration in the Air}
Physics of Atomic Nuclei 66 (2003) 462; http://dx.doi.org/10.1134/1.1563705.

\bibitem[{Gavriljuk et~al.(2011) Gavriljuk, Gangapshev, Kuzminov, Panasenko and Ratkevich}]{Gavr11}
Ju.M.~Gavriljuk, A.M.~Gangapshev, V.V.~Kuzminov, S.I.~Panasenko, S.S.~Ratkevich,
    \emph{Monitoring the $^{222}$Rn Concentration in the Air of Low-Background Laboratories by Means of an Ion-Pulse Ionization Chamber.}
    Bulletin of the Russian Academy of Sciences: Physics 75 (2011) 547; 
    http://dx.doi.org/10.3103/S1062873811040150.

\bibitem[{Gavriljuk et~al.(2009) Gavriljuk, Gangapshev, Kuzminov, Panasenko and Ratkevich}]{Gavr09}
Ju.M.~Gavriljuk, A.M.~Gangapshev, V.V.~Kazalov, V.V.~Kuzminov, S.I.~Panasenko, S.S.~Ratkevich,
    \emph{An Ion Pulse Ionization Chamber for Spectrometric Measurements of Low Surface a Activities.}
    Instruments and Experimental Techniques 52 (2009) 173; http://dx.doi.org/10.1134/S0020441209020055.

\bibitem[{A.~Savitzky and M.J.E.~Golay (1964) Savitzky and Golay}]{Savitzky64}
A.~Savitzky and M.J.E.~Golay,
        \emph{Smoothing and Differentiation of Data by Simplified Least Squares Procedures.}
        {Analytical Chemistry} 36 (1964) 1627. http://dx.doi.org/10.1021/ac60214a047.

\bibitem[{Kotrappa (1988) Kotrappa}]{Kotrappa88}
P.~Kotrappa, J.C.Dempsey, J.B.~Hickey, L.B.~Stieff,
\emph{An Electret Passive Environmental 222Rn Monitor Based on Ionization Measurement.}
Health Physics. 54(1) (1988) 47.

\bibitem[{Agostinelli et~al.(2003)}]{Agostinelli2003}
    Agostinelli~S., et~al., 
    \emph{GEANT4-a simulation toolkit.}
    Nuclear Instruments and Methods A 506 (2003) 250-303;
    http://dx.doi.org/10.1016/S0168-9002(03)01368-8

\bibitem{Lindhard63}
J.~Lindhard, V.~Nielsen, M.~Scharff, and P.V.~Thomsen, K.Dan.~Vidensk,
Selsk. Mat. Fys. Medd. 33,  (1963) 10;
available at http://www.sdu.dk/Bibliotek/matfys.

\bibitem[{Raizer (1980) Raizer}]{Raizer80}
Y.P.~Raizer,
\emph{Fundamentals of Modern Physics discharge processes.} - Moscow: Atomizdat, 1980.

\bibitem[{Gledenov (1993) Gledenov}] {Gledenov93}
Y.M.~Gledenov, V.I.~Salatski, P.V.~Sedyshev,
\emph{The $^{14}$N(n,p)$^{14}$C reaction cross section for thermal neutrons}
Zeitschrift f\"{u}r Physik A Hadrons and Nuclei V.346(4), (1993) 307;
http://dx.doi.org/10.1007/BF01292522.

\bibitem[{Alekseenko (2011) Alekseenko}] {Alekseenko1302}
V.V.~Alekseenko, Ju.M.~Gavriljuk, V.V.~Kuzminov and S.S.~Ratkevich
\emph{The results of measurements helium proportional counter CH-04 neutron background on objects BNO INR.}
Preprint INR RAS, 1302/2011, Moscow, 2011.

\bibitem[{Alekseenko (2011) Alekseenko}] {Alekseenko1303}
V.V.~Alekseenko, Ju.M.~Gavriljuk, V.V.~Kuzminov,
\emph{Features characteristics scintillation detector of thermal neutrons [ZnS(Ag) +6LiF] in different measurement conditions.}
Preprint INR RAS, 1303/2011, Moscow, 2011.


\bibitem{Siegban65}
{Alpha-, Beta- and Gamma-Ray Spectroscopy}, edited by K.Siegban,
North-Holland Publishing Company, Amsterdam, 1965.

\bibitem{Grupen96}
C.~Grupen, {Particle Detectors},
Cambridge, University Press, 1996.

\bibitem{Fano}
U.~Fano,
\emph{Ionization Yield of Radiations. II. The Fluctuation of the Number of Ions},
Phys.Rev. 72 (1947) 26;
http://dx.doi.org/10.1103/PhysRev.72.26.

\bibitem{Kiko}
J.Kiko,
\emph{Detector for $^{222}$Rn measurements in air at the 1 mBq/m$^3$ level.}
Nuclear Instruments and Methods A 460 (2001) 272;
http://dx.doi.org/10.1016/S0168-9002(00)01082-2.

\end{thebibliography}
\end{document}